\documentclass{aa}  

\usepackage{graphicx}
\usepackage{txfonts}
\usepackage{color}

\newcommand{\om}[1]{{ #1}}

\begin{document} 
   \title{The abundance of dwarf galaxies around low-mass giants in the Local Volume}
   \author{Oliver M\"uller
          \inst{1}
          \and
          Helmut Jerjen\inst{2}
          }

 \institute{Observatoire Astronomique de Strasbourg  (ObAS),
Universite de Strasbourg - CNRS, UMR 7550 Strasbourg, France\\
 \email{oliver.muller@astro.unistra.fr}
\and
 Research School of Astronomy and Astrophysics, Australian National University, Canberra,
ACT 2611, Australia
}

   \date{Received tba; accepted tba}

 
  \abstract
   {The abundance of satellite dwarf galaxies has long been considered a crucial test for the current model of cosmology leading to the well-known missing satellite problem. Recent advances in both simulations and observations have allowed to study dwarf galaxies around host galaxies in more detail. We have surveyed a 72 deg$^2$ area of the nearby Sculptor group using the Dark Energy Camera -- also encompassing the two low-mass Local Volume galaxies NGC\,24 and NGC\,45 residing behind the group -- to search for hitherto undetected dwarf galaxies.  Apart from the previously known dwarf galaxies we have found only two new candidates down to a $3\sigma$ surface brightness detection limit of 27.4\,$r$\,mag arcsec$^{-2}$. Both systems are in projection close to NGC\,24. \om{However, one of these candidates could be an  ultra-diffuse galaxy associated to a background galaxy.} We compared the number of known dwarf galaxy candidates around NGC\,24, NGC\,45, and five other well-studied low-mass giant galaxies (NGC\,1156, NGC\,2403, NGC\,5023, M\,33, and the LMC) with predictions from cosmological simulations and found that for the stellar-to-halo mass models considered, the observed satellite numbers tend to be on the lower end of the expected range. This could either mean that there is an over-prediction of luminous subhalos in $\Lambda$CDM or \om{-- and more likely --}  that we are missing some of the satellite members due to observational biases.}
   \maketitle
%

\section{Introduction}
In the standard cosmological framework dwarf galaxies are thought to be the building blocks of the visible Universe. Larger galaxies form through a cascade of minor mergers of these dwarf galaxies \citep{2012AnP...524..507F} leading to  a strong correlation between the number of dwarf galaxy satellites and the mass of the host galaxy \citep[e.g., ][]{2019ApJ...870...50J}, as the more massive halos are able to accrete more matter. The left-overs of these accretion processes are still detectable today as satellite galaxies swarming the central galaxies.
The specific frequency of satellites and their luminosities  can be described by the galaxy luminosity function (LF, \citealt{1988ARA&A..26..509B}) and be compared to cosmological predictions. This led to the well-known missing satellite \citep{1999ApJ...524L..19M} and the too-big-to-fail \citep{2011MNRAS.415L..40B} problems. These challenges for the $\Lambda$CDM model of cosmology, however, have mainly been studied in the Local Group of galaxies. Only recently new technology allowed to survey for dwarf galaxies in other nearby groups to a sufficient surface brightness depth that such cosmological tests can be conducted more systematically.

The Local Volume ($D<11$\,Mpc, \citealt{1979AN....300..181K,2004AJ....127.2031K,2013AJ....145..101K}) hosts over 30 large galaxies with total luminosities in excess of $M_{tot}\approx -20$\,$K$ mag. Several surveys have targeted the more prominent of these giant galaxies like  M\,83 \citep{2015A&A...583A..79M}, Centaurus\,A \citep{2014ApJ...795L..35C,2016ApJ...823...19C,2017A&A...597A...7M,2018ApJ...867L..15T}, and others \citep{2014ApJ...787L..37M,2015AstBu..70..379K,2016A&A...588A..89J,2017ApJ...848...19P,2018ApJ...863..152S,2019ApJ...885..153B,2020ApJ...891..144C,2020arXiv200308352D}. Furthermore, dwarf galaxy surveys reach to even more distant galaxy clusters \citep{2017A&A...608A.142V,2019ApJS..245...10W}, groups \citep{2017ApJ...847....4G,2018ApJ...868...96C,2020MNRAS.491.1901H} and the field \citep{2018ApJ...857..104G,2019MNRAS.488.2143P}.

These surveys have revealed first interesting results. For instance, the recent deep search for dwarf galaxies around the giant spiral galaxy M\,94 (NGC 4736; $M_*\approx 4\times 10^{10} M_\sun$, \citealt{2013AJ....145..101K}; $D=4.2$\,Mpc, \citealt{2011ApJS..195...18R}) has detected only two satellites \citep{2018ApJ...863..152S}. Its LF seems to be inconsistent with predictions from high-resolution dark matter simulations, having far too few satellites for such a massive galaxy. In addition to that, \citet{2019ApJ...885..153B} argued for a strong variation of the LF derived from the environments of the giant galaxies in the Local Volume. This scatter seems to be larger than what is expected from the concordance model (but see also \citealt{2020ApJ...891..144C} for a different view). On the other hand, there seems to emerge a correlation between the number of satellites and the bulge-to-disk ratio of the host galaxies, which is again unexpected in $\Lambda$CDM \citep{2019ApJ...870...50J,2020MNRAS.493L..44J}. All these findings ask for more observations of different environments. While there appears to be a consensus that the missing-satellite problem for the Milky Way \om{and the Andromeda galaxy are} resolved today \citep{2007ApJ...670..313S,2016MNRAS.457.1931S,2018MNRAS.478..548S}, there are still discrepancies between observations and the $\Lambda$CDM model when it comes to the abundance of dwarf galaxies around the giants in the Local Volume. 

One of the closest galaxy aggregates from our point of view is the Sculptor group in the southern hemisphere. It is long known to be a loose association of galaxies \citep{1998AJ....116.2873J,2000AJ....119..593J}, stretching approximately from NGC\,55 (D=2.13\,Mpc, \citealt{2005A&A...431..127T}) to NGC\,59 (4.89\,
Mpc, \citealt{2013AJ....146...86T}), with the starburst spiral galaxy NGC\,253 at a distance of 3.7\,Mpc \citep{2013AJ....145..101K,2015MNRAS.450.3935L} being the dominant member. An imaging survey of its immediate surroundings has revealed only two faint companions  \citep{2014ApJ...793L...7S,2016ApJ...816L...5T}, which already suggests that indeed the Sculptor group is a low-density environment. This is surprising because NGC\,253 is as massive \citep{2015MNRAS.450.3935L} as the Milky Way or the Andromeda galaxy, which both host a plethora of dwarf galaxies \citep[e.g., ][]{2010ApJ...712L.103B,2011ApJ...732...76R,2015ApJ...804L..44K}. Due to its proximity to us, the Sculptor group covers several hundred square degrees in the sky, making a full coverage of the group observationally highly demanding. Here we present the results from a first 72 deg$^2$ survey based on dedicated CCD observations of the eastern part of the Sculptor group, which encompasses two additional giant galaxies, NGC\,24 (D=7.67\,Mpc, \citealt{2013AJ....146...86T}) and NGC\,45 (D=6.67\,Mpc, \citealt{2013AJ....146...86T}), which are behind the group but still in the Local Volume. 

This paper is structured as follows: in Section \ref{sec:obs} we present the observations and data reduction, in Section \ref{sec:search} we discuss the visual and automated search for dwarf galaxies and perform photometry, in Section \ref{sec:member} we argue the membership of the newly found dwarf galaxy candidates, in Section \ref{sec:lf} we compare the luminosity function of seven low-mass host galaxies to predictions from dark matter simulations, and finally in Section \ref{sec:sum} we draw our conclusions.

\section{Observation and data reduction}
\label{sec:obs}
\begin{figure}[ht]
\includegraphics[width=\linewidth]{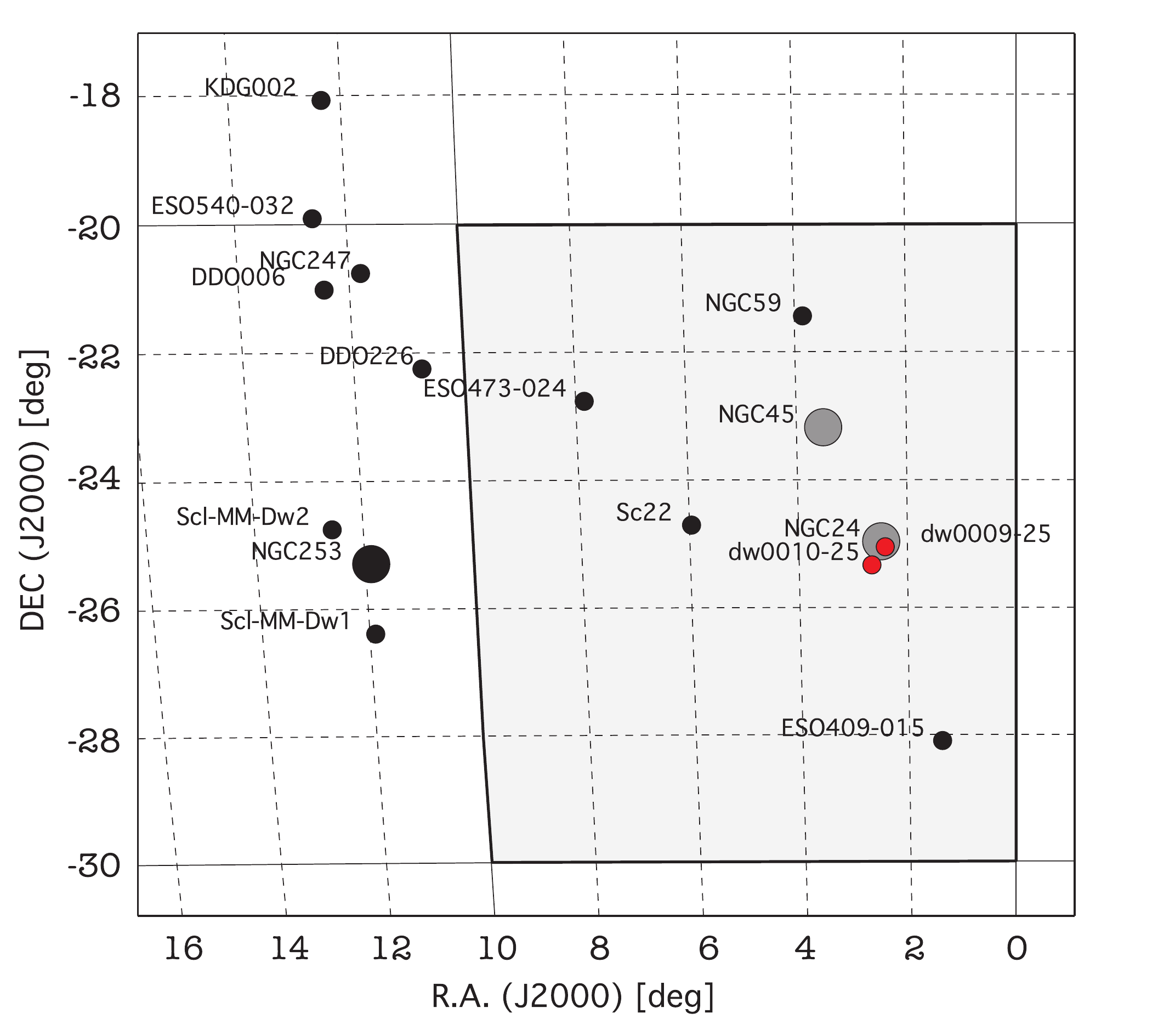}
\caption{The field of the Sculptor group. The small black dots correspond to dwarf galaxies in the Sculptor group with its main galaxy NGC\,253 (large black dot). The small gray dots are background dwarf galaxies, which are still within the Local Volume ($D<11$\,Mpc. The two large gray dots are the two LSB giants NGC\,24 and NGC\,45. The red dots are our dwarf galaxy candidates dw0009-25 and dw0010-25. The grey area corresponds to our survey footprint.}
\label{field}
\end{figure}

\om{
The $gr$ CCD images for our search of Sculptor group dwarf candidates were obtained as part of the Stromlo Milky Way Satellite Survey (e.g. \citealt{2015ApJ...803...63K,2016ApJ...820..119K}). Imaging data were collected for a total of $\sim$ 500\,deg$^2$ with the DECam at the 4\,m Blanco telescope at CTIO over three photometric nights from 17th to 19th July 2014. DECam is an array of sixty-two 2k$\times$4k CCD detectors with a 2.2 deg$^2$ field of view and a pixel scale of $0\farcs263$(unbinned). We obtained a series of $4 \times 60$\,s dithered exposures in the $g$ and $r$ band for each pointing under photometric conditions. To cover the 72\,deg$^2$ of the Sculptor region we acquired a total of 24 pointings (see Fig.\,\ref{field} for the survey footprint). The seeing in those Sculptor fields ranged from $0\farcs79\leq\sigma_g\leq 1\farcs22$ and $0\farcs80\leq \sigma_r\leq 1\farcs16$, with median values $\mu_{1/2}(\sigma_g)=0\farcs94$ and $\mu_{1/2}(\sigma_r) =0\farcs94$,  respectively.
%
 
The images were reduced via the DECam community pipeline \citep{2014ASPC..485..379V}, which included overscan subtraction, bias calibration, flat field gain calibration, single exposure cosmic ray masking, illumination correction, astrometric calibration to refine the world coordinate system of each frame, and photometric calibration.}
For the sky subtraction we employed {\it SEP} -- the python version \citep{2016JOSS....1...58B} of Source Extractor \citep{1996A&AS..117..393B}. A 400 $\times$ 400 px$^2$ box was used as reference for the local background measurement. This size is large enough ($\approx$ 2\,kpc $\times$ 2\,kpc) to avoid over-subtraction of the sky on the scales we are interested in, i.e. the typical angular size of a dwarf galaxy at the distance of the Sculptor group.  After the individual pointings were background subtracted, the two bands were combined using the {\it SWarp} program  \citep{2002ASPC..281..228B} to make a final, deep image. 

The photometric zero points of the $g$ and $r$ images were derived using the PanSTARRS DR2 catalog \citep{2016arXiv161205242M}. We performed aperture photometry with the python package {\it photutils} \citep{Bradley_2019_2533376} and compared them to standard stars in the magnitude range 15.0 - 19.5\,mag. A linear regression with a 0.3\,mag clipping was applied to derive the slope and intercept in each band and for each field. The slope was consistent with 1.0 so that the intercept corresponds to the zero point. \om{The errors are 0.04\,mag in the $g$-band and 0.08\,mag in the $r$-band.}

\section{Search for new dwarf galaxies in the Scl Group region}
\label{sec:search}
In our survey area only four Local Volume dwarf galaxies were known to date: ESO\,409-15 ($D=8.7$\,Mpc; \citealt{2013AJ....146...86T}), NGC\,59 ($D=4.9$; \citealt{2013AJ....146...86T}), Sculptor-dE1 (\citealt{2000AJ....119..593J}, $D=4.2$\,Mpc \citealt{2003A&A...404...93K}), and ESO\,473-24 ($D=9.9$\,Mpc; \citealt{2013AJ....145..101K}). Additionally, two Local Volume spiral galaxies reside in the field: NGC\,24 ($D=7.7$\,Mpc; \citealt{2013AJ....146...86T}), an edge-on Sc galaxy, and NGC\,45 ($D=6.7$\,Mpc;  \citealt{2013AJ....146...86T}), a low-surface brightness Sdm galaxy. These two latter galaxies are of low mass, with total baryonic plus dark matter mass estimates of $2.8\times10^{10}$\,M$_\odot$ and $3.7\times10^{10}$\,M$_\odot$, respectively \citep{2006AJ....132.2527C}.

To find new dwarf galaxies we rely on both visual inspection of the images and on an automatic detection pipeline. The search is done on the stacked $gr$ images.

\subsection{Visual inspection}
\begin{figure}[ht]
\centering
\includegraphics[width=0.49\linewidth]{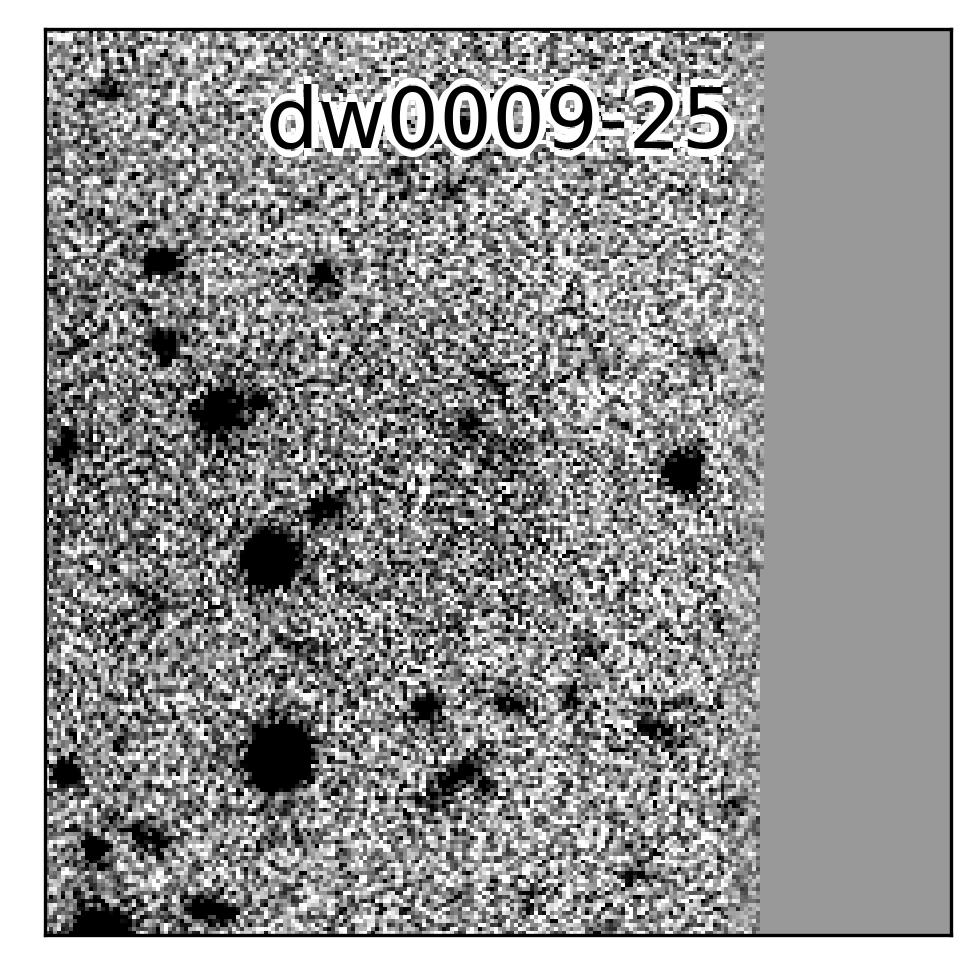}
\includegraphics[width=0.49\linewidth]{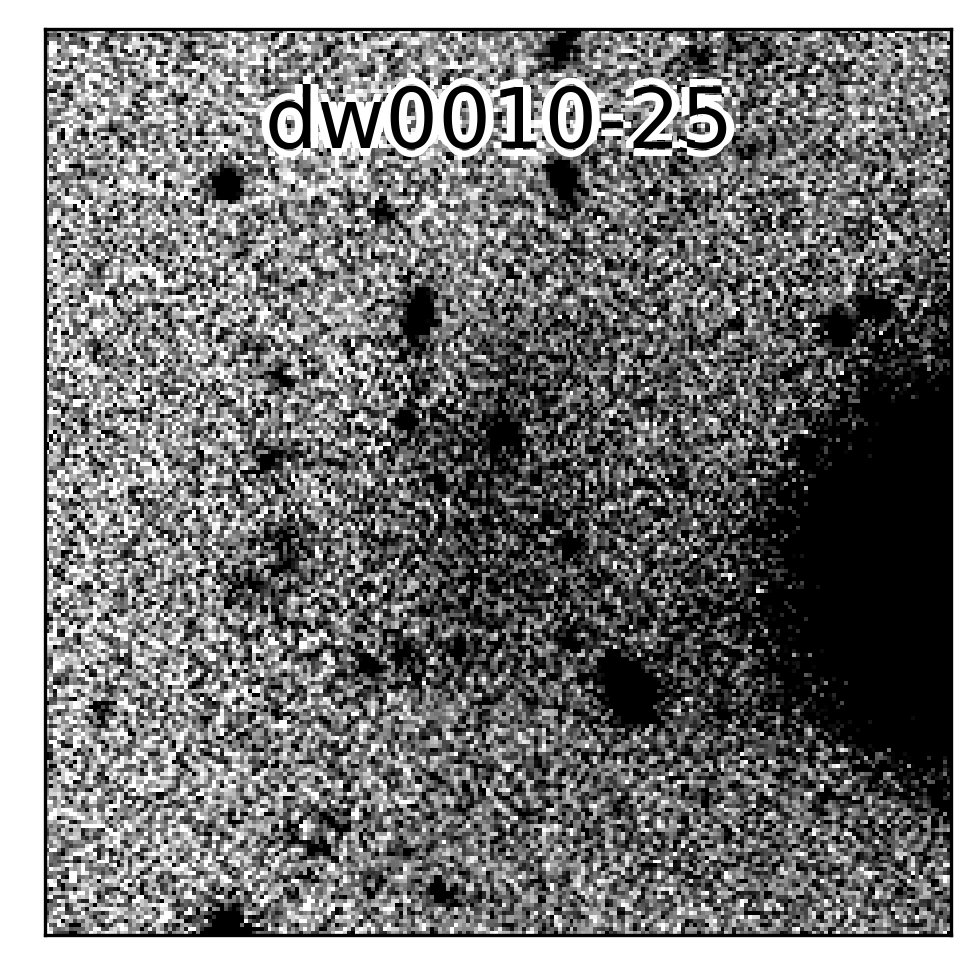}
\caption{The stacked $gr$-band images of the newly discovered dwarf galaxy candidates dw0009-25 (left) and dw0010-25 (right). North is to the top, east to the left, the vertical length of the images corresponds to one arcmin. We note that dw0009-25 was detected close to the CCD edge.}
\label{dwarf}
\end{figure}

The classical approach to search for faint low-surface brightness galaxies is via careful visual inspection of the digital images. This approach has been proven successful to detect dozens of new dwarf galaxies in surveys around other nearby galaxies \citep[e.g., ][]{2000AJ....119..593J,2000A&AS..145..415K,2002MNRAS.335..712T,2009AJ....137.3009C,2014ApJ...787L..37M,2017ApJ...848...19P,2018A&A...615A.105M,2020MNRAS.491.1901H,2020ApJ...891...18B}. To enhance any low-surface brightness features, \om{a standard image processing algorithm is applied to the images, i.e. the convolution with a Gaussian kernel (with $\sigma$ of 1.5\,px).}  Afterwards all images are scanned by eye for diffuse and extended patches which resemble the morphology of dwarf galaxies. This approach leads to a good understanding of the quality and limitations of the data at hand, which will also benefit the more automated detection techniques. Assuming that a faint dwarf galaxy detectable in our images has a typical effective radius of $r_{\rm eff}=140$\,pc \citep{2019A&A...629A..18M} this would give an angular size of 30\,px at the mean distance of the Sculptor Group (3.45\,Mpc) and 14\,px at the distance of NGC\,24. Such an object is well distinguishable from larger background dwarf galaxies, the latter being more compact compared to their surface brightness. Our visual search revealed one good dwarf galaxy candidate which we dubbed dw0010-25, and another potential candidate named dw009-25 (see Fig.\,\ref{dwarf}).

\subsection{Automated detection with MTObjects}
Novel development of automated detection algorithms show promising results. One of the best software available today is MTObjects \citep{teeninga2013bi,teeninga2015improved}, a max-tree based algorithm, which was developed for medical image analysis and re-written for astronomical purposes (see e.g. \citealt{2019MNRAS.488.2143P}). It is non-parametric, this is, it automatically searches for the best parameters for a given image, making MTObjects straightforward to use. MTObjects provides a segmentation map and a catalog of basic photometric parameters for each detection. On these preliminary catalogs we apply further quality cuts for size, total magnitude and surface brightness. The limits for the effective radius $r_{eff}$ was set to be  0.1-3.0\,kpc at the adopted distance of the Sculptor group (3.7\,Mpc),
This reduced the number of objects to a total of 3372 for the whole survey area. 
We inspected each of the detections by eye and only kept the good dwarf galaxy candidates leaving us with only one object, the already known Sculptor-dE1 dwarf. This means the visually detected dwarf galaxy candidate dw0010-25 was not picked up by MTObjects. We attribute this to the presence of a background spiral galaxy close in its proximity (see Fig.\,\ref{dwarf}). The other dwarf galaxy candidate dw0009-25 was too faint to get detected by MTObjects (see next paragraph).
The large number of false positives were mainly from artefacts at the edges of the CCD images. 

\begin{figure}[ht]
\centering
\includegraphics[width=\linewidth]{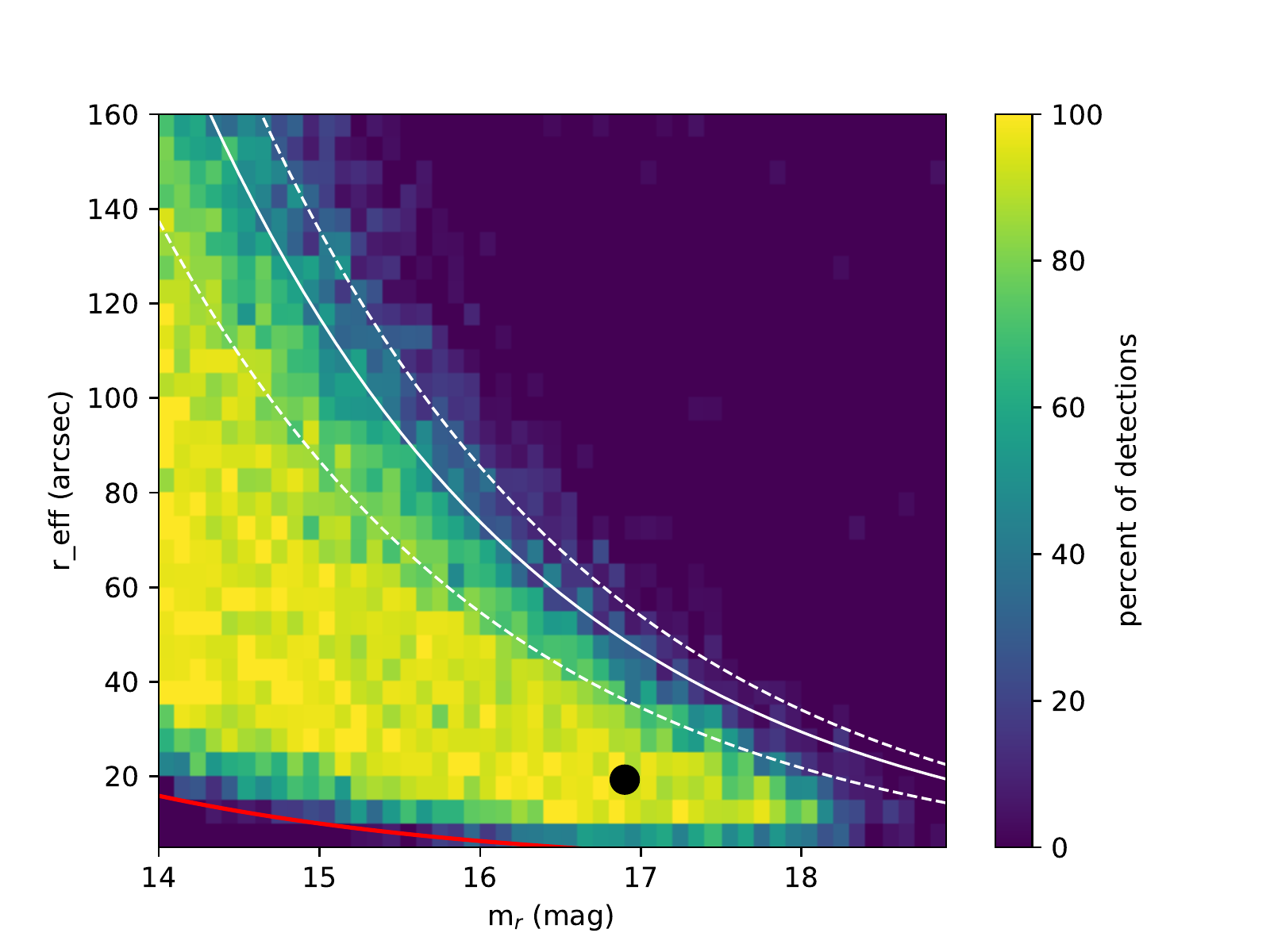}
\caption{Result from the artificial dwarf galaxy test with MTObjects showing the recovery rate (color scheme) as a function of total apparent magnitude $m_r$ and half-light radius $r_{eff}$. \om{The  red line indicates the lower surface brightness cut (22\,mag arcesc$^{-2}$), the  white line the estimated completeness at 50\% (27.5\,mag arcesc$^{-2}$), and the dashed white lines the 75\% ($26.7$\,mag  arcsec$^{-2}$) and 25\% ($27.7$\,mag  arcsec$^{-2}$) limit, respectively. The black dot indicates the known dwarf galaxy Sculptor-dE1.}}
\label{detection}
\end{figure}

To assess the completeness of the dwarf galaxy detection, we have injected artificial dwarf galaxies into the images and re-run MTObjects with the same quality cuts.  \om{The dwarfs were modelled with a S\'ersic profile \citep{1968adga.book.....S}, employing different apparent magnitudes (i.e. between 14\,mag and 19\,mag) and effective radii (i.e. between 5\,arcsec and 160\,arcsec). S\'ersic indices between $n=0.5$ and $n=1.5$ were chosen to model the dwarfs, with randomly drawn ellipticities between $e=0.0$ and $e=0.6$.} Each dwarf was separated by 1000 pixels and was arranged in an equidistant grid. The separation of 1000 pixels corresponds to roughly two times the maximal allowed effective radius in our quality cuts. This makes sure that there is no issue of overcrowding the field. By design, some of the artificial dwarf galaxies will be at the edge of the image, such that we also probe any bias coming from dwarf galaxies being at the edge of the camera. Ultimately,
the detection rate was derived from a comparison between the input and detection list, with an error tolerance of one effective radius for the position of the dwarf. \om{This process was repeated across different fields with a total of 8000 artificial dwarf galaxies per field, where the position of the grid was randomized during each iteration. From the detection rate, we derived the mean surface brightness limit by binning the magnitudes and effective radii like in Figure\,\ref{detection} with bin sizes of 0.2\,mag and 10\,arcsec, respectively. Then we extracted the bins as data points where we achieved a detection rate between 80 to 70 percent for the 75\% completeness limit, data points between 45 to 55 percent for the 50\% completeness limit, and data points between 20 to 30 percent for the 25\% completeness limit. Through these data points we fitted the formula
\begin{equation}
    r_{\rm  eff}^2= 10^{0.4(\mu_{\rm lim}-m_r-0.753)} / \pi,
\end{equation}
where $\mu_{\rm lim}$ is the quantity we are solving for.
We have performed this limit estimation on six different fields and find  50\% completeness limits of: $27.19\pm0.10$\,mag  arcsec$^{-2}$, $27.55\pm0.09$\,mag  arcsec$^{-2}$, $27.17\pm0.09$\,mag  arcsec$^{-2}$, $27.09\pm0.12$\,mag  arcsec$^{-2}$,  $27.38\pm0.12$\,mag  arcsec$^{-2}$, and  $27.07\pm0.09$\,mag  arcsec$^{-2}$. To estimate the overall completeness limit of the survey, we have combined these individual runs to one final diagram with bin sizes of 0.1\,mag and 5\,arcsec, respectively. A mean surface brightness limit $\mu_{\rm lim}$ of $27.34\pm0.08$\,mag  arcsec$^{-2}$ in the $r$-band best describes our 50\% detection rate limit ($26.68\pm0.10$\,mag  arcsec$^{-2}$ at  75\% and $27.65\pm0.08$\,mag  arcsec$^{-2}$ at  25\%). The results are illustrated in Figure\,\ref{detection}.} This is consistent with the numbers estimated in \citet{2015A&A...583A..79M}, which was part of the same imaging campaign with the same observation strategy. For the apparent magnitude $m_r$ (x-axis in Figure\,\ref{detection}), the recovery rate drops bellow 50\% at 18.2\,mag  in the $r$-band, which translates into a completeness limit \om{for the automated detection of dwarf galaxies} of $M_r=-9.6$\,mag at the distance of NGC\,253, $-11.1$\,mag at the distance of NGC\,24, and  $-10.9$\,mag at the distance of NGC\,45, respectively. The faintest detectable objects have a total apparent $r$ magnitude of $\sim18.5$. This limit is brighter than one of the dwarf candidate we detected by eye ($m_r=20.6$\,mag) \om{by two magnitudes}, meaning that at the faint end of the luminosity function the  visual inspection performs better than the automated detection pipeline with MTObjects.

We also looked into the question how much of the survey area is covered by extended objects such as the main galaxies, background galaxies and bright/saturated stars which could potentially hide dwarf galaxies. To estimate a number we have used the parameters given by MTObjects. Namely we integrated the area of all objects with effective radii larger than 7.5\,px. To do this, we have approximated the area of each object fulfilling the criteria with a circle with two times the effective radius. This gives a coverage of 0.2\% which can potentially obfuscate the dwarf galaxies. This is indeed negligible and will not significantly alter the abundance of dwarf galaxies.

\subsection{Photometry}
The photometry for the two dwarf galaxy candidates was done using GALFIT \citep{2002AJ....124..266P}. To achieve the best S\'ersic fit we provided  GALFIT with a mask created from the segmentation map of MTObjects, which we further adjusted by hand. Due to their low surface brightness levels we had to impose tighter constraints on the center of the object (i.e. $\pm$5\,px from our  initial guess based on the overall surface brightness distribution of the object). We performed photometry in both the $g$ and $r$ bands. \om{The errors for the magnitudes are a combination of the uncertainty from the photometric zero point and the error from GALFIT, the rest is taken from the uncertainties provided by GALFIT.}
The measured photometric parameters are given in Table\,\ref{properties}. We have measured the surface brightness limit around our dwarf galaxy candidates in randomly distributed $10\times10$\,arcsec boxes \citep{2019A&A...624L...6M} and derived a $3\sigma$ limit of 27.4\,mag arcsec$^{-2}$  \om{in the $r$-band}. \om{This limit is well consistent with the 50\% detection limit we derived from our artificial galaxy tests.}

\begin{table}[!htb]
\caption{Photometric properties of the dwarf galaxy candidates.}
\centering                          
\begin{tabular}{l c c }        
\hline\hline                 
& dw0009-25 & dw0010-25  \\    
\hline      \\[-2mm]                  
R.A. (J2000) & 00:09:37.8 & 00:10:38.2 \vspace{1mm}\\ 
DEC (J2000) & $-$25:02:57.2 & $-$25:20:05.8 \vspace{1mm}\\
$m_g$ (mag)& 21.50$\pm$0.50 & 18.04$\pm$0.13 \vspace{1mm}\\
$m_r$ (mag)& 20.62$\pm$0.11 & $17.47\pm0.08$ \vspace{1mm}\\
$A_g$, $A_r$ (mag)& 0.06, 0.04 & 0.06, 0.04\vspace{1mm}\\
$(g-r)_0$ (mag) & 0.87$\pm$0.51 & 0.55$\pm$0.15 \vspace{1mm}\\
$r_{eff,r}$ (arcsec)& 8.0$\pm$0.9 & 30.2$\pm$1.4 \vspace{1mm}\\
$\mu_{eff,r}$ (mag arcsec$^{-2}$)& 26.9$\pm$0.2 & 26.7$\pm$0.1 \vspace{1mm}\\
$PA$ (north to east)& 1$\pm$19 & 168$\pm$4\vspace{1mm}\\
$e=1-b/a$ & 0.16$\pm$0.07 & 0.19$\pm$0.02\vspace{1mm}\\
S\'ersic index $n$& 0.75$\pm$0.10 & 1.33$\pm$0.05\\
\hline
\end{tabular}
\label{properties}
\end{table}

\section{Membership of the dwarf galaxy candidate}
\label{sec:member}

Both dwarf galaxy candidates dw0009-25 and dw0010-25 are in projection close to the edge-on spiral galaxy NGC\,24 and also close to each other (see Fig.\,\ref{field}). Their on-sky separation is only 0.11\,deg and 0.40\,deg from NGC\,24, which at its distance ($D=7.3$\,Mpc) corresponds to 14.5\,kpc and 52.5\,kpc, respectively. Assuming that the two dwarfs are satellites of NGC\,24, the effective radius and absolute magnitude would be $r_{eff}=283$\,pc and $M_r=-8.7$\,mag for dw0009-25 and $r_{eff}=1069$\,pc and  $M_r=-11.9$\,mag for dw0010-25. The next nearest luminous galaxy is NGC\,45, which has an on-sky separation to the dwarfs of more than 2.3\,deg, corresponding to a distance larger than 240\,kpc at the distance of $D=6.6$\,Mpc for both, which is comparable to the virial radius and thus a rather large distance for any satellite galaxy.
There is a remote possibility that the dwarf candidates are associated to NGC\,253 ($D=3.7$\,Mpc). They have an on-sky separation of \textasciitilde8.3\,deg, which translates into linear distances of \textasciitilde550\,kpc. The physical properties of the dwarf galaxy candidate then would boil down to $M_r=-7.2$\,mag and  $r_{eff}=144$\,pc for dw0009-25 and   $M_r=-10.4$\,mag and  $r_{eff}=542$\,pc for dw0010-25. 

How do these quantities compare to the structural parameters of other known dwarf galaxies? 
\citet{2020MNRAS.491.1901H} studied a population of over 2000 dwarf galaxies in the nearby universe and presented the structural parameters in their Figure 11. Our dwarf galaxy candidates are compatible with these dwarf galaxies at either distance, so no potential membership can be confirmed or excluded. 
Also the integrated $(g-r)_0$ color of 0.87\,mag and 0.55\,mag are consistent with the mean color of 0.46$\pm$0.26\,mag of dwarfs in the Cen\,A group \citep{2018A&A...615A.105M}, being part of the same survey. Due to their proximity we assign dw0009-25 and dw0010-25 to  NGC\,24 until follow-up distance measurements can confirm their true association. 

There is one caveat, dw0010-25 being very close to the interacting background spiral LEDA\,133604 (see Fig.\,\ref{dwarf}), which is at a redshift of $z=0.065$ \citep{1996ApJS..107..201L}, i.e., $\approx$290\,Mpc away from us. At that distance dw0010-25 would have an absolute magnitude of $M_r=-19.8$\,mag and an effective radius of $r_{eff}=42$\,kpc, which seems highly unrealistic. The same is true for dw0009-25. It has the background galaxy LEDA\,783199 in its vicinity, but no distance information is available in that case. Assuming a conservative distance of 100\,Mpc this would translate into  $M_r=-14.4$\,mag and an effective radius of $r_{eff}=4$\,kpc, which are plausible properties for an ultra-diffuse galaxy in the background \citep[e.g. ][]{2020ApJS..247...46B}.


\section{The luminosity function of low-mass giant galaxies}
\label{sec:lf}

\begin{figure}[ht]
\includegraphics[width=\linewidth]{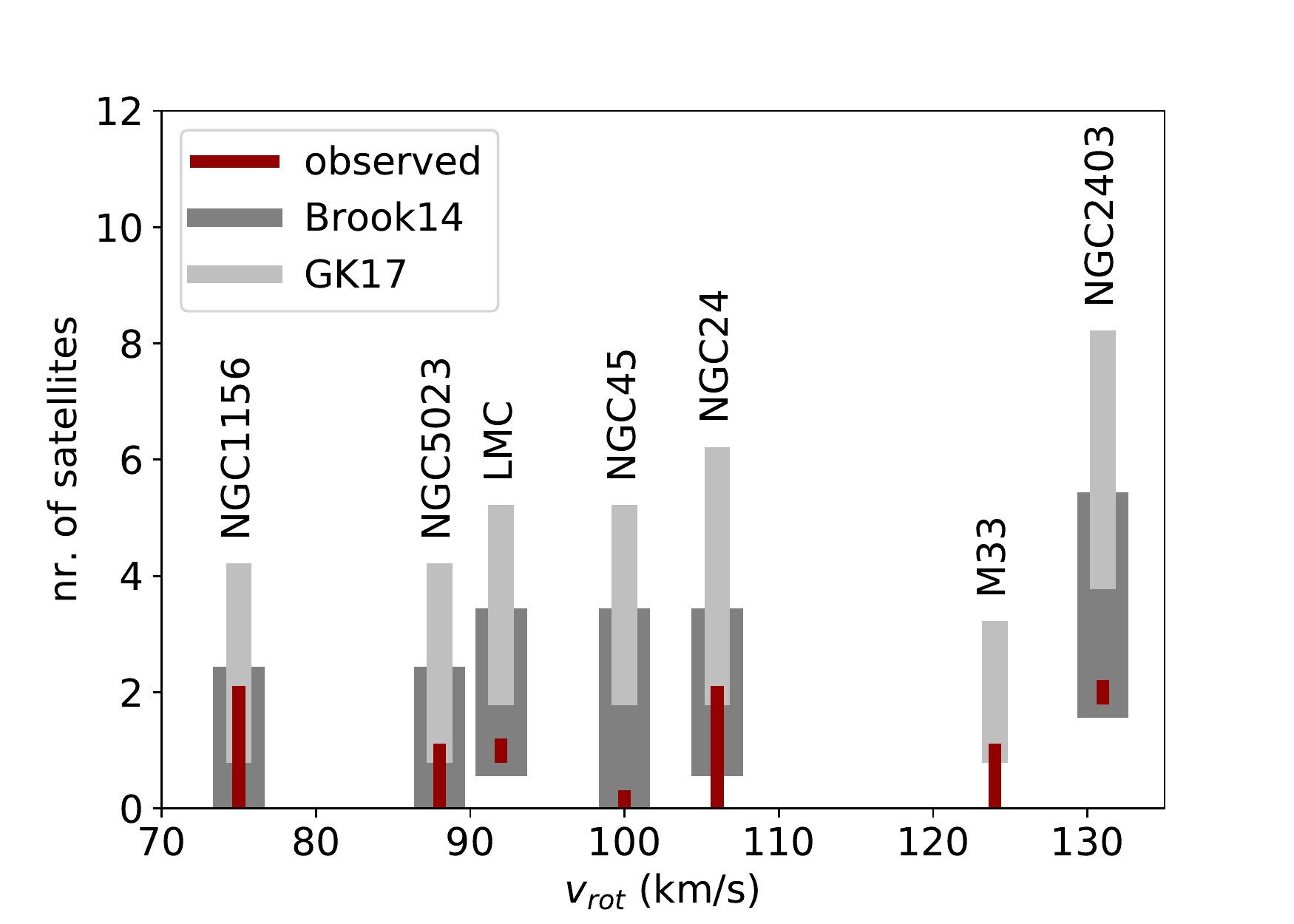}
\caption{The predicted (gray boxes) and observed (red lines) number of satellites as a function of the rotation curve of the host galaxy. For the predictions, the assumed stellar-to-halo mass model by \citet{2014ApJ...784L..14B} and \citet{2017MNRAS.471.1709G} in \citet{2017MNRAS.472.1060D} are employed.}
\label{nr_sat}
\end{figure}

All dwarf galaxy satellites of the two host galaxies NGC\,24 and NGC\,45 should have been identified within the $M_V<-10$\,mag luminosity range  by now. For NGC\,24, we find two possible satellites while there are none for NGC\,45. Is this expected? Indeed, with total masses of $2.8\times10^{10}$\,M$_\odot$ and $3.7\times10^{10}$\,M$_\odot$ \citep{2006AJ....132.2527C}, respectively, the two host galaxies are at the low-mass end of the giant galaxy population. Let us compare these results to other such galaxies. The LMC is probably the most iconic low-mass giant with an enclosed mass of  $1.7\times10^{10}$\,M$_\odot$ \citep{2014ApJ...781..121V}. It has one satellite -- the SMC -- more massive than $10^5$\,M$_\odot$, but has also many light-weight satellites \citep{2020MNRAS.495.2554E} which would be impossible to detect outside of the Local Group. \citet{2020arXiv200602443C} conducted a deep survey for dwarf galaxies around 10 giants within the Local Volume and studied their luminosity functions. Two of those, NGC\,1156 and NGC\,5023 are of similar low mass than the two giants studied here. NGC\,1156 has two possible dwarf galaxy satellites and one is known for NGC\,5023. 
Another well studied low-mass giant galaxy is M\,33 with an observed dark matter mass of $5\times10^{10}$\,M$_\odot$ \citep{2003MNRAS.342..199C}. Only one possible satellite has been reported for M\,33 to date \citep{2009ApJ...705..758M}. The MADCASH survey \citep{2016ApJ...828L...5C} aimed at finding satellite companions of LMC analogs. In a first search around NGC\,2403 one dwarf galaxy was discovered, increasing the number of known NGC\,2403 satellites to two.
The flat part of the rotation curves of all these low-mass giant galaxies reaches between $v_{rot,flat}=75$ and 131 km\,s$^{-1}$, which sets them apart from galaxies like the Milky Way \citep{1989ApJ...342..272F} or  NGC\,253 \citep{2015MNRAS.450.3935L}, with significantly higher rotation velocities in the range between $200<v_{rot,flat}<220$ km\,s$^{-1}$.
We have compiled the values for these low-mass giant galaxies in Table\,\ref{giants}.

\begin{table*}[!htb]
\caption{Low-mass giant galaxies and their number of satellite galaxies {within the virial radius.}}
\centering                          
\begin{tabular}{l c c r r r r}        
\hline\hline                 
& $M_K$ & $M_{*,host}$ & $v_{rot, flat}$  & observed & predicted (Brook14) & predicted (GK17) \\ 
Name & (mag) &  ($10^9 M_\odot$) & (km\,s$^{-1}$) & ($M_{tot}>10^5\,M_\odot$) & ($M_{tot}>10^5\,M_\odot$) & ($M_{tot}>10^5\,M_\odot$) \\ 
\hline      \\[-2mm]                  
NGC\,24 & $-20.3$ & $1.6$& 106$\pm$8$^{1}$ &  0-2 & 1-3 & 2-6 \\  
NGC\,45 & $-20.0$ & $1.2$& 100$\pm$1$^{2}$ &  0 & 0-3 & 2-5 \\  
NGC\,1156$^{*}$ & $-20.0$ & $1.2$ &75$\pm$4$^{3}$ &  0-2 & 0-2 & 1-4\\ 
NGC\,2403 & $-21.3$ & $4.1$& 131$\pm$5$^{4}$ & 2 & 2-5 & 4-8\\
NGC\,5023 & $-19.2$ & $0.6$& 88$\pm$5$^{5}$&  0-1 & 0-2 & 1-4\\  
M\,33$^{+}$& $-20.8$ & $2.6$& 124$\pm$11$^{6}$ &  0-1 & --- & 1-3  \\
LMC & $-20.2$ & $1.5$ &  92$\pm$19$^{7}$ &  1 & 0-2 & 1-4\\  
\hline
\end{tabular}
\tablefoot{The $M_K$ magnitudes were taken from the Local Volume catalog \citep{2013AJ....145..101K}. The baryonic mass $M_{*,host}$  of the galaxy is derived from $M_K$ and a mass-to-light ratio of 0.6 \citep{2017ApJ...836..152L} and a solar $K$ band luminosity of $K=3.28$\,mag \citep{1998gaas.book.....B}.
The flat rotation velocity $v_{rot,flat}$ corresponds to the flat part of the host galaxy rotation curve. The references for $v_{rot,flat}$ are:
(1) \citet{2008MNRAS.385..553D};
(2) \citet{2006AJ....132.2527C};
(3) \citet{1990A&AS...86....1K};
(4) \citet{2006MNRAS.367..469D};
(5) \citet{2013MNRAS.434.2069K};
(6) \citet{2014A&A...572A..23C};
(7) \citet{2014ApJ...781..121V}. We have calculated the mean value of the flat part of the rotation curve ourselves when not explicitly given in these references. \om{The observed number of satellites must be considered a lower limit.}
The predicted number of satellites comes from the assumed stellar-to-halo mass model relation by Brook14 \citep{2014ApJ...784L..14B} and GK17 \citep{2017MNRAS.471.1709G} in \citet{2017MNRAS.472.1060D} for dwarf galaxies more massive than $10^5$\,M$_\odot$. (*) For NGC\,1156 no $\Lambda$CDM prediction of the number of satellites is available. Here we have assumed the same number as for NGC\,5023 due to their similar $v_{rot,flat}$. (+) For M\,33 we use the prediction by \citet{2018MNRAS.480.1883P} who adapted the \citet{2017MNRAS.472.1060D} predictions for the LMC with a GK17 model. }
\label{giants}
\end{table*}

How do the observed number of possible dwarf satellites compare with predictions from the current $\Lambda$CDM model of cosmology? \citet{2017MNRAS.472.1060D} has calculated the expected number of satellites with masses larger than $10^5$\,M$_\odot$. This mass limit coincides with the depths reached in the various surveys allowing for a \om{more or less} direct comparison. \om{More or less, because the completeness to this limit is still not fully reached, underestimating the total abundance of observed satellite galaxies.} 
In Table\,\ref{giants} and Fig.\,\ref{nr_sat} we present the number of expected and observed satellites for the stellar-to-halo mass models by \citet{2014ApJ...784L..14B} and \citet{2017MNRAS.471.1709G}. They correspond to the low and high end of the estimates. For the Brook model, the number of possible dwarf galaxies coincide for all of the seven here studied systems with the 80\% confidence interval of \citet{2017MNRAS.472.1060D}.
However, there is a systematic trend. The observed satellite number for most of the low-mass host galaxies is at the lower end of the $\Lambda$CDM predictions. The difference is even more evident when comparing to the predictions from the GK17 model. Half of the systems \om{seems off from the} observations. This can be interpreted that either $\Lambda$CDM is over-producing massive subhalos, or \om{-- more likely --} that some dwarf satellites have been overlooked in all these surveys. The former is a known problem in simulations \citep{2014MNRAS.438.2578G},  while the latter is observationally unavoidable, with some dwarf galaxy satellites potentially being obscured by the host galaxy or other sources (i.e. cirrus, bright stars, large background galaxies, or noise), \om{or simply too faint to be detectable}. This incompleteness becomes an even larger problem towards the faint end of the luminosity function where larger numbers of smaller dwarf galaxies are expected. For NGC\,24, the automated detection limit is at $-10.9$\,mag, which translates into a stellar luminosity of $17\times10^{5}$\,L$_\odot$ \om{(with an absolute $r$-band luminosity of the sun of 4.61, \citealt{2018ApJS..236...47W}).} This is still a factor ten higher than what was used in the predictions by \citet{2017MNRAS.472.1060D} -- note though that the dwarf galaxy candidate dw0009-25 has a luminosity of $-8.7$\,mag at the distance of NGC\,24\om{, corresponding to $2\times10^{5}$\,L$_\odot$, so being right in the ballpark of the luminosity limit.} For a statistically  robust comparison the luminosity function should be complete down to $\approx -8$\,mag \om{ or below}. \om{Complete in the sense that both the depth and the recovery rate are of high enough quality to be certain that all satellites were observed.}
\om{So the most natural explanation from any deviation from the predictions is}  that the currently missing satellites are in this incomplete luminosity range. \om{ However, for NGC\,2403 a sufficient depth was reached, but the field of view was too small to cover its virial radius and thus the survey lacks the full coverage of the satellite system.} There could be a handful of dwarf galaxies still residing outside the survey footprint \citep{2016ApJ...828L...5C}.

\section{Summary and conclusion}
\label{sec:sum}
The Sculptor group is the closest galaxy aggregate to the Milky Way and thus extends over a large area of the sky. With the Dark Energy Camera and in the $g$ and $r$ bands, we have observed a 72 deg$^2$ area in the Eastern part of the Sculptor group. This region also encloses the two low-mass spiral galaxies NGC\,24 and NGC\,45 of the Local Volume, which are  further in the background and therefore unrelated to the Sculptor group. We have searched for low-surface brightness dwarf galaxies employing two different strategies. We have visually inspected all CCD images by eye and we have applied automatic detection methods to find the dwarf galaxies. To test the level of completeness for the search we have injected artificial dwarf galaxies and found to be 50\% complete down to 18.2 mag
or $M_r\approx -11$ to $-10$\,mag. In our visual search we have found only two dwarf galaxy candidate, which are in projection closest to NGC\,24. These candidate were not picked up by the automated detection algorithm due to the proximity to a background galaxy in one case, and its faint luminosity in the other. The latter case is noteworthy, as it shows that a visual search for dwarf galaxies can reach some deeper limit than the automated one. 
The physical properties of the candidates are compatible with the scaling relations defined by known dwarf galaxies \citep{2020MNRAS.491.1901H}, independent of the association to the two background Local Volume galaxies or the Sculptor group. Follow-up observations are needed to pin down their true host galaxy memberships via distance measurements \citep{2019ApJ...879...13C,2019A&A...629L...2M,2019ApJ...880L..11M}. 

We have compared the observed number of dwarf galaxy candidates around NGC\,24 and NGC\,45, as  well as for five other well-studied low-mass giant galaxies (NGC\,1156, NGC\,2403, NGC\,5023, M\,33, and the LMC), to predictions from high-resolution dark matter simulation employing different stellar-to-halo mass models. We find that for the mass model by \citet{2014ApJ...784L..14B} the observed abundance of dwarf galaxies generally follows the predictions, but still being systematically on the lower end. As we are dealing with very low number statistics the difference could either arise from observational biases or be first evidence for a true tension between the observed and predicted satellite population for low-mass galaxies. For the \citet{2017MNRAS.471.1709G} model there seems to be an over-prediction of satellites and the tension becomes more evident.  The strongest disagreement is found for the most luminous sample galaxy NGC\,2403.  It will be interesting to see whether this trend persists once the environments of more low-mass giants have been carefully investigated, e.g. by the MADCASH survey \citep{2016ApJ...828L...5C}, \om{or it is indeed rather an observational bias due to the incompleteness kicking in at the low-end of the luminosity function. On this aspect it is interesting that a recent Hubble Space Telescope study of low-mass galaxies at redshifts 0.1 to 0.8 has found good agreement between the observed luminosity function  down to $M_V=-15$\,mag and the predicted abundance of satellites \citep{2020arXiv200805479R}. Furthermore, \citet{2020arXiv200602443C} found that the abundance of satellites of more massive Local Volume galaxies is in good agreement between the observed and expected number of satellites. This is compatible with our earlier assessment of the Cen\,A group, where we found that the luminosity function matches the prediction within the 90\% confidence interval \citep{2019A&A...629A..18M}. This indicates that in general, the luminosity function is well described by our current model of cosmology.}



\begin{acknowledgements}
We thank the referee for the constructive report, which helped to clarify and improve the manuscript.
O.M. wants to thank the Swiss National Science Foundation for financial support. H.J. acknowledges support from the Australian Research Council through the Discovery Project DP150100862.
\end{acknowledgements}

   \bibliographystyle{aa}
\bibliography{aanda}

\end{document}